\documentclass[journal]{IEEEtran}

\usepackage{amsmath}
\usepackage{amssymb}
\usepackage{amsfonts}
\usepackage{graphicx}
\usepackage{float}
\usepackage[noadjust]{cite}
\usepackage{subfig}
\usepackage{wrapfig}

\newcommand{\R}{{\mathbb{R}}}

\newtheorem{theorem}{Theorem}[section]

\newtheorem{problem}[theorem]{Problem}
\newtheorem{remark}[theorem]{Remark}
\newtheorem{assumption}[theorem]{Assumption}

\def\L2{{\cal L}_2}
\def\L2e{{\cal L}_{2e}}

\def\diag{\mbox{diag}}

\newcommand{\mc}{\mathcal}

\def\begequarr{\begin{eqnarray}}
\def\endequarr{\end{eqnarray}}
\def\begequarrs{\begin{eqnarray}}
\def\endequarrs{\end{eqnarray}}
\def\begarr{\begin{array}}
\def\endarr{\end{array}}
\def\begequ{\begin{equation}}
\def\endequ{\end{equation}}
\def\lab{\label}
\def\begdes{\begin{description}}
\def\enddes{\end{description}}
\def\begenu{\begin{enumerate}}
\def\begite{\begin{itemize}}
\def\endite{\end{itemize}}
\def\endenu{\end{enumerate}}

\def\lef[{\left[\begin{array}}
\def\rig]{\end{array}\right]}

\def\begcen{\begin{center}}
\def\endcen{\end{center}}
\def\begrem{\begin{remark}\rm}
\def\endrem{\end{remark}}

\usepackage{color}

\newcommand{\jo}[1]{\textcolor{black}{#1}}

\title{Online Estimation of Power System Inertia Using Dynamic Regressor Extension and Mixing}
\author{Johannes Schiffer, Petros Aristidou and Romeo Ortega
\thanks{J. Schiffer is with Brandenburgische Technische Universit\"at Cottbus-Senftenberg, 03046 Cottbus, Germany (e-mail: schiffer@b-tu.de).}
\thanks{P. Aristidou is with University of Leeds, LS2 9JT Leeds, UK (e-mail: p.aristidou@leeds.ac.uk).}
\thanks{R. Ortega is with CNRS L2S, 91192 Gif Sur Yvette, France (e-mail: romeo.ortega@lss.supelec.fr).}
}

\begin{document}
\maketitle

\begin{abstract}
The increasing penetration of power-electronic-interfaced devices is expected to have a significant effect on the overall system inertia and a crucial impact on the system dynamics. In the future, the reduction of inertia will have drastic consequences on protection and real-time control and will play a crucial role in the system operation. Therefore, in a highly deregulated and uncertain environment, it is necessary for Transmission System Operators to be able to monitor the system inertia in real time. We address this problem by developing and validating an online inertia estimation algorithm. The estimator is derived using the recently proposed dynamic regressor and mixing procedure. The performance of the estimator is demonstrated via several test cases using the 1013-machine ENTSO-E dynamic model.
\end{abstract}


\section{Introduction}

\subsection{Motivation and Existing Literature} 
\IEEEPARstart{T}{raditionally}, power systems have been relying on the inertia provided by synchronous generators to provide the necessary energy buffer for smoothing out sudden power imbalances (deficit or surplus) in the system. The inertia of conventional generators creates a direct physical connection to the grid, thus providing instantaneous power when necessary and helping to curb the frequency deviations created by abrupt power imbalances.

In modern power systems, conventional power plants are gradually being replaced by power-electronic(PE)-interfaced generators (mainly, integrating renewable energy sources) and high capacity network interconnections being implemented through high voltage direct current (HVDC) links. Consequently, the replacement of synchronous generators with PE-interfaced devices decreases the available inertia in the system and can lead to much faster frequency dynamics in the grid \cite{ulbig14,winter15,orum15,milano18}. In this situation, the dynamic behavior can endanger the system by stressing the control and protection schemes, which were not designed to operate in such conditions \cite{ulbig14,winter15}, leading to cascaded failures and disconnections. Moreover, the remaining legacy components could be endangered if they cannot withstand the emerging dynamics~\cite{ulbig14,milano18}.

In addition, in the present-day deregulated and uncertain environment, it becomes very difficult for Transmission System Operators (TSOs) to accurately track the system inertia and provide guarantees about the system stability \cite{ulbig14,orum15}. This inability leads to overly conservative operational planning scenarios, which inflate the operational costs. Hence, the capability to monitor the inertia available in the system in real time would allow TSOs to operate with lower security margins (and cost) by taking appropriate actions to secure the system operation. \jo{Moreover, inertia-related constraints can put stress on the power markets, thus increasing the cost for other power market actors~\cite{Davarinejad17}.}

Several techniques for power system inertia estimation have been proposed in the literature. The majority of available inertia estimators only work offline, {\em i.e.,} with data collected after an event \cite{inoue97,chassin05,ashton15,zografos17,zografos17_2,zografos18} or over a certain time window \cite{tuttelberg18}, and are based on a simplified swing equation model.
While such {\em a posteriori} inertia estimation can be very useful, the information might arrive too late for any preventive or corrective actions to take place. In \cite{ashton15}, phasor-measurement-unit (PMU) measurements are used to estimate regional inertia values and to then reconstruct the total inertia. The estimation relies on the accurate detection of a suitable event and requires post-processing the PMU data.

A near real-time, iterative, inertia estimation algorithm combining Least-Squares, Newton-Raphson and Modal Assurance Criterion techniques is proposed in \cite{guo12}. In \cite{wall12,wall14} an online estimation algorithm is presented based on the linearized swing equation and a set of four filters implemented as sliding data windows. Yet, in addition to the active power flow also the rate of change, {\em i.e.,} the derivative, of the frequency at the generators needs to be known. Furthermore, the estimation method is only applicable immediately after a disturbance and critically depends on the exact knowledge of the time at which the disturbance occurs. To ensure the latter an additional disturbance time estimation algorithm is proposed in \cite{wall14}. A statistical approach using steady-state and relatively small frequency variations is presented in \cite{cao16}. The proposed online-estimation method in \cite{zhang17} requires the injection of an additional probing signal, which complicates its implementation. \jo{Finally, a simple method employed by some TSOs is based on the monitoring of the circuit-breaker status of synchronous generators~\cite{orum15} and knowledge of each generator inertia. Although simple, this approach requires real-time monitoring of all synchronous generators in the system and an accurate knowledge of the generator parameters.}

\subsection{Contributions}
Our main contributions in this work are three-fold:
\begenu
\item We propose an algorithm which allows to estimate in real time the inertia constant and the aggregated mechanical power setpoint of a large-scale power system. The algorithm is derived using a first-order nonlinear aggregated power system model in combination with the recently proposed dynamic regressor and mixing (DREM) procedure~\cite{aranovskiy16}, \jo{which already has been applied very successfully to a variety of electrical engineering applications \cite{aranovskiy15_2, bobtsov15,bobtsov17}.}
\item The performance of the estimator is demonstrated on a 1013-machine ENTSO-E test system with 21382 buses and 133997 states, which is implemented in the dynamic simulation software RAMSES~\cite{ALC2016ja}. 
\jo{As with any dynamic parameter estimation method \cite{sastry11,astolfi07,narendra12}, also with DREM a sufficiently large system excitation is required for an accurate estimation \cite{aranovskiy16}. The considered scenarios for this purpose investigated in the paper consist of 25 generator outages and a rescheduling event.} Neither the location nor the size of the perturbations need to be known and in all scenarios the same estimator gains are used. Furthermore, our proposed algorithm only requires frequency and electrical power measurements from primary-frequency-controlled (PFC) generators, which typically represent a subset of all machines in the system. As pointed out in \cite{wall12}, such data can be provided using synchronized measurement technology (SMT)~\cite{terzija11}.
\item We show that the PFC power injection signal can be well-approximated by a simple, aggregated model of the turbine-governor dynamics of the PFC units. Naturally, this leads to reduction of the required measurements and we illustrate that this approximation
only results in a minor reduction of the achievable estimator accuracy. 
\endenu

The remainder of the paper is structured as follows. The aggregated power system model and the problem statement are introduced in Section~\ref{sec:model}. The DREM-based inertia estimator is derived in Section~\ref{sec:regest}. The employed aggregated power system model is validated in Section~\ref{sec:entsoe} using simulation results from a detailed dynamic 1013-machine ENTSO-E system obtained with the software RAMSES. The estimator is tested via a nominal outage scenario and the performance is investigated further in Section~\ref{sec:test} with \jo{24} additional cases. Final conclusions and a brief outlook on future work are given in Section~\ref{sec:con}.

\section{Aggregated Power System Model and Problem Formulation}
\label{sec:model}
\subsection{Center of Inertia Frequency Dynamics} 
\label{sec:coi}
For the purpose of deriving an online inertia estimator, we seek to represent the frequency dynamics of a primary-controlled power system using an equivalent reduced-order model. We assume that $N_\text{PFC}>0$ is the number of PFC generators in the system and $N_\text{unc}\geq0$ is the number of \jo{rotational} generators (and large motors) without PFC. Thus, $N=N_\text{PFC}+N_\text{unc}$ is the total number of \jo{rotational} generators (and large motors) in the system. 
 It is well-known \cite{anderson12,weckesser17,orum15} that the principal frequency dynamics of a power system can be described by the evolution of the center of inertia (COI) speed, which is defined as
\begequ
\omega_\text{COI}=\frac{\sum_{i=1}^N H_i\omega_i}{\sum_{i=1}^N H_i}.
\endequ 

Here, $\omega_i:\R_{\geq0}\to\R_{>0}$ is the angular frequency of the rotor of the $i$-th unit and $H_i\in\R_{>0}$ the $i$-th unit's inertia constant\footnote{The sets $\R_{\geq0}$ and $\R_{>0}$ denote the non-negative and positive real numbers, respectively. Hence, the notation $\omega_i:\R_{\geq0}\to\R_{>0}$ means that $\omega_i(t)$ is a function, which takes positive values for all times $t\in\R_{\geq0}.$}.

Let $S_{B_i}\in\R_{>0}$ denote the power rating of the $i$-th unit and $P_{m_i}\in\R$ its scheduled (constant) mechanical power. Then
$$S_B=\sum_{i=1}^N S_{B_i}$$ 
represents the total power rating of the considered system. Furthermore, the total inertia constant of the power system is given by
\begequ
H_\text{tot}=\frac{\sum_{i=1}^N H_i S_{B_i}}{\sum_{i=1}^N S_{B_i}}
\notag 
\endequ 
Likewise, the total mechanical power is given by
\begequ
P_{m,\text{tot}}=P_{m,\text{PFC}}+P_{m,\text{unc}}=\sum_{i=1}^{N_\text{PFC}} P_{m_i}+\sum_{i=1}^{N_\text{unc}} P_{m_i}.
\label{Pmtot}
\endequ

Let $P_{e,\text{PFC}}:\R_{\geq0}\to \R_{\geq 0}$ denote the aggregated electrical power by the PFC generators and $P_{e,\text{unc}}:\R_{\geq0}\to \R_{\geq 0}$ that by the non-PFC units.
Then, the total generated electrical power is denoted by $P_{e,\text{tot}}:\R_{\geq0}\to \R_{\geq 0}$ and satisfies 
\begequ
P_{e,\text{tot}}=P_{e,\text{PFC}}+P_{e,\text{unc}}=P_\text{load}+P_\text{loss}-P_\text{ren},
\label{Petot}
\endequ 
where $P_{\text{load}}:\R_{\geq0}\to \R_{\geq0}$ is the total load demand in the system, $P_{\text{loss}}:\R_{\geq0}\to \R_{\geq0}$ are the total losses and $P_{\text{ren}}:\R_{\geq0}\to \R_{\geq0}$ is the total renewable generation power.

With these considerations, the principal frequency dynamics of the system can be described by the aggregated swing equation \cite{anderson12}
\begequ
\begin{split} 
\dot \omega_\text{COI} &=\frac{\omega_0^2}{2 H_\text{tot} S_B}\frac{\Delta P}{\omega_\text{COI}}=\frac{\omega_0^2}{2 H_\text{tot} S_B}\left(\frac{P_{m,\text{tot}}+ P_\text{PFC,tot}-P_{e,\text{tot}}}{\omega_\text{COI}}\right),
\end{split}
\lab{coi}
\endequ 
where $\omega_0\in\R_{>0}$ is the nominal network frequency and $P_\text{PFC,tot}:\R_{\geq0}\to\R_{\geq0}$ is the total \jo{mechanical power injection due to PFC action}\footnote{For the purpose of inertia estimation, we are interested in fast time scales. Therefore secondary control actions are neglected in the torque balance \eqref{coi}.}. 

\subsection{Problem Formulation}
Determining directly the variables $\omega_\text{COI}$, $P_{m,\text{tot}}$ and $P_{e,\text{tot}}$ in \eqref{coi} is very hard in practice. Hence, any real-time inertia estimator based on \eqref{coi} would not be implementable. However, modern power systems are equipped with advanced SMTs or wide-area measurement systems (WAMS) \cite{terzija11} that can provide a TSO access to the measurements of the PFC units. 
Using these measurements, we can approximate equation \eqref{coi} describing the COI frequency dynamics by expressing the power balance $\Delta P$ on the right hand-side with information from the PFC generators, {\em i.e.,} 
\begequ
\begin{split} 
\Delta P=	P_{m,\text{PFC}}+P_\text{PFC,tot}-P_{e,\text{PFC}},
\end{split}
\notag 
\endequ 
where from \eqref{Petot}
\begequ
P_{e,\text{PFC}}=-P_{e,\text{unc}}+P_\text{load}+P_\text{loss}-P_\text{ren}.
\endequ 
In addition, the COI frequency is approximated by the average frequency of the PFC units, {\em i.e.,}
\begequ
\omega_\text{av}=\frac{\sum_{i=1}^{N_\text{PFC}} \omega_i}{N_\text{PFC}}.
\notag 
\endequ 
These steps yield the following approximation of the COI frequency dynamics \eqref{coi}
\begequ
\begin{split} 
	\dot \omega_\text{av} &=\frac{\omega_0^2}{2 H_\text{tot} S_B}\left(\frac{P_{m,\text{PFC}}+ P_\text{PFC,tot}-P_{e,\text{PFC}}}{\omega_\text{av}}\right),
\end{split}
\lab{sweq}
\endequ 
which is used in the remainder of this work.

Furthermore, the above discussion on the available measurements is summarized in the following assumption.

\begin{assumption}
	The signals $\omega_\text{av}$, $P_\text{PFC,tot}$ and $P_{e,\text{PFC}}$ are measurable.
	\lab{ass1}
\end{assumption}

If information on the type and parameters of the turbine-governor systems is available, $P_\text{PFC,tot}$ can be obtained from the average frequency $\omega_\text{av}$. Then, Assumption~\ref{ass1} can be further relaxed. This is demonstrated in Section~\ref{sec:entsoe} for the ENTSO-E 1013-machine system.

\jo{In addition to the online application, {\em i.e.}, using data arriving in real time, the proposed algorithm can also be used offline with the (standard) requirement being that all data is time stamped. This, however, would result in a delayed estimation in the sense that it no longer is executed in real time.}

We are interested in the following problem.

\begin{problem}
Consider the system \eqref{sweq} with measurable signals  $\omega_\text{av}$, $P_\text{PFC,tot}$ and $P_{e,\text{PFC}}$.
Identify the parameters $H_\text{tot}$ and $P_{m,\text{PFC}}$.
\lab{prob1}
\end{problem} 

We remark that, as in \eqref{coi}, the parameter $H_\text{tot}$ is the {\em total} inertia constant of PFC and non-PFC units. Hence, we seek to estimate the total system inertia using only partial system information from the PFC units. In addition, the total mechanical power $P_{m,\text{PFC}}$ should be estimated. This is useful in case any frequency variations are caused by changes in $P_{m,\text{PFC}}$, {\em e.g.}, due to rescheduling or an outage of a PFC generator.

\section{System Parametrization, Regression Construction and DREM Estimator}
\label{sec:regest} 
\subsection{System Parametrization}
We begin our exposition by bringing the model \eqref{sweq} in the standard form for dynamic parameter estimation schemes \cite{sastry11,astolfi07,narendra12}. First, we define the new variables
\begequ
y=\omega_\text{av},\quad
x=P_\text{PFC,tot},\quad u=P_{e,\text{PFC}},
\notag 
\endequ
the constant
\begequ
b_1=\frac{\omega_0^2}{2S_B},
\notag 
\endequ
and the parameters
\begequ 
\eta_1=\frac{1}{H_\text{tot}},\quad \eta_2=\frac{P_{m,\text{PFC}}}{H_\text{tot}}.
\label{eta}
\endequ 
Then, \eqref{sweq} takes the form
\begequ
\begin{split} 
	\dot y &=\eta_1 b_1\left(\frac{x-u}{y}\right)+\eta_2 \frac{b_1}{y}.
\end{split}
\label{y_dot}
\endequ
This system parametrization is used for the regression construction and estimator development detailed below. 

\subsection{Regression Construction}
To address Problem~\ref{prob1}, we construct a regression by using the dynamics described by \eqref{y_dot}. Let $$p=\frac{d}{dt}$$ denote a differentiation operator.
Then, by applying the operator $\frac{\alpha}{(p+\alpha)}$ with some $\alpha>0$ to \eqref{y_dot}, we obtain
\begequ
\frac{\alpha p}{(p+\alpha)}y=\eta_1\frac{\alpha}{(p+\alpha)}b_1\left(\frac{x-u}{y}\right)+\eta_2 \frac{\alpha}{(p+\alpha)}\frac{b_1}{y}+\epsilon,
\label{y_dot2}
\endequ 
where $\epsilon$ is an exponentially decaying term stemming from the filters' initial conditions.
Then, by setting
\begequ
\begin{split} 
\lab{xi12}
\xi_1&=\frac{\alpha p}{(p+\alpha)}y,\,
\xi_2=\frac{\alpha}{(p+\alpha)}b_1\left(\frac{x-u}{y}\right),\,
\xi_3=\frac{\alpha}{(p+\alpha)}\frac{b_1}{y},
\end{split}
\notag
\endequ 
we can rewrite \eqref{y_dot2} compactly as
\jo{ 
\begequ
\xi_1=\eta_1\xi_2+\eta_2\xi_3+\epsilon,
\notag 
\endequ
}
or, equivalently, 
\begequ
z=\phi^\top \eta+\epsilon,
\lab{reg}
\endequ 
with
\begequ
\begin{split}
z&=\xi_1,\quad 
\phi^\top=\begin{bmatrix} \xi_2 & \xi_3\end{bmatrix},\quad \eta^\top=\begin{bmatrix} \eta_1 & \eta_2 \end{bmatrix}.
\end{split}
\notag
\endequ	 

\subsection{DREM Parameter Estimator}
To construct a parameter estimator for the regressor equation \eqref{reg}, we follow the DREM procedure \cite{aranovskiy16}. The regression \eqref{reg} is of dimension $1$, but the number of unknown parameters is $q=2$. Hence, the first step of the procedure is to introduce a linear, $\mc L_\infty$-stable operator $\mc H:\mc L_\infty\to\mc L_\infty$, the output of which may be decomposed for any bounded input as
\begequ
(\cdot)_f(t)=[\mathcal H(\cdot)](t)+\epsilon_t,
\notag 
\endequ 
where $\epsilon_t$ is an exponentially decaying term. This operator can be chosen in several ways. For instance, a possible choice would be an exponentially stable linear time-invariant (LTI) filter of the form
$$
\mc H(p)=\frac{\alpha}{p+\alpha},\quad \alpha\neq0.
$$
Another option is to choose a delay operator, {\em i.e.,}
\begequ 
[\mc H(\cdot)](t)=(\cdot)(t-d),
\label{opdelay}
\endequ 
for some $d>0$. The impact of different operators on the transient performance of the DREM estimator is extensively discussed in \cite{aranovskiy16,ortega18}. In the authors' experience the delay operator \eqref{opdelay} has proven very successful in power engineering applications. Therefore, this option is also chosen for the present application.

In the second step, we apply the operator \eqref{opdelay} to the regressor in \eqref{reg}. This yields the filtered regression\footnote{To simplify the presentation in the sequel we neglect the $\epsilon$ and $\epsilon_t$ terms, see also \cite{aranovskiy16}.}
\begequ
z_f=\phi_f^\top \eta,
\label{regfilt}
\endequ 
which in the present case is equivalent to
\begequ 
z(t-d)=\phi^\top(t-d) \eta.
\label{zdelay}
\endequ 
\jo{Many standard software packages, such as Matlab/Simulink, contain delay operators, which can be used to implement \eqref{zdelay} in a straighforward manner.}

The third step in DREM consists of piling up the original regression \eqref{reg} with the filtered regression \eqref{regfilt}, which gives the extended regression system
\begequ
\begin{bmatrix}
z\\z_f
\end{bmatrix}=\Phi \eta,
\label{regext}
\endequ 
where
\begequ
\Phi=\begin{bmatrix} \phi^\top \\\phi_f^\top \end{bmatrix}.
\label{Phi}
\endequ 
Next, we have that
$$
\text{adj}(\Phi)\Phi=\Delta I_2,
$$
where $I_2$ denotes the $2\times 2 $ identity matrix and
\begequ
\Delta=\det(\Phi).
\label{Delta}
\endequ 
Consequently, by premultiplying \eqref{regext} with the adjunct matrix of $\Phi$ we obtain two scalar regression equations
\begequ
\mc Z=\begin{bmatrix}\mc Z_1\\\mc Z_2 \end{bmatrix}=\Delta\eta,
\lab{regdrem}
\endequ 
where
\begequ
\mc Z = \text{adj}(\Phi)\begin{bmatrix}
z\\z_f
\end{bmatrix}.
\notag 
\endequ 

\begin{figure}
	\includegraphics[width=\linewidth]{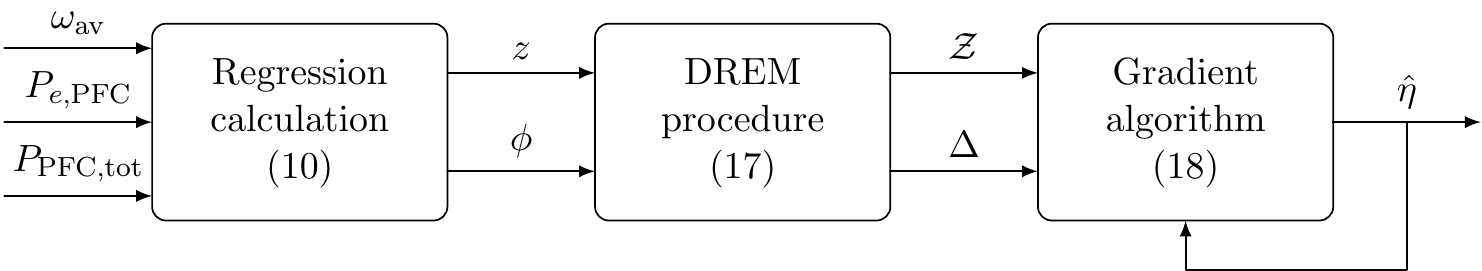}
	\caption{Block diagram of the DREM-based online inertia estimator.}
	\label{fig:drem_est}
\end{figure}

We define $\hat \eta$ as the estimated value of the parameter vector $\eta$. By using the decoupled regression \eqref{regdrem}, we can then estimate $\eta$ via the gradient algorithm \cite{aranovskiy15}
\begequ
\begin{split} 
\dot {\hat \eta}_1&=
\gamma_1 \Delta(\mc Z_1-\Delta \hat \eta_1),\\ 
\dot {\hat \eta}_2&=\gamma_2 \Delta (\mc Z_2-\Delta \hat \eta_2),
\end{split} 
\label{inest}
\endequ 
where $\gamma_1>0$ and $\gamma_2>0$ are tuning gains. A block-diagram of the proposed estimator is shown in Fig.~\ref{fig:drem_est}. 

Recall that a signal $\xi:\R_{\geq0} \to \R^m$ is in $\mc L_2$ if its $\mc L_2$-norm $\|\xi\|_{\mc L_2}$, given by
$$
\|\xi\|_{\mc L_{2}}=\sqrt{\int_0^\infty \xi^\top (t) \xi(t)d t},
$$
is finite.
\jo{Introducing the error coordinates $\tilde \eta=\hat\eta-\eta$, recalling the fact that $\eta$ is a vector of constant parameters and using \eqref{regdrem}, the error dynamics corresponding to \eqref{inest} are given by
\begequ
\begin{split} 
\dot {\tilde \eta}=\dot {\hat \eta}&=\diag(\gamma_1,\gamma_2)\Delta (\mc Z-\Delta \hat \eta)\\
&=\diag(\gamma_1,\gamma_2)\Delta (\Delta\eta-\Delta \hat \eta)=-\diag(\gamma_1,\gamma_2)\Delta^2 \tilde\eta,
\end{split} 
\label{inesterr}
\endequ 
where $\diag(\cdot)$ denotes a diagonal matrix.} Hence, we see that the following equivalence holds
\begequ
\lim_{t\to\infty} \tilde \eta=0 \quad \Leftrightarrow \quad \Delta \notin \mc L_2.
\label{deltanotl2}
\endequ 
The requirement \eqref{deltanotl2} is different from that in conventional parameter identification techniques. The usual persistency of excitation (PE) condition is defined as \cite{sastry11,astolfi07,narendra12}
\begequ
\int_t^{t+\tau} \phi(s)\phi^\top (s)ds\geq \delta I_2,
\notag 
\endequ 
for some $\tau>0$ and $\delta.$ Hence, PE is a property of the regressor $\phi,$ while DREM requires the determinant of the matrix $\Phi$ not to be square integrable. 
We refer to \cite{aranovskiy16,ortega18} for an in-depth analysis of the convergence and robustness properties of DREM parameter estimators as well as for examples of regressors, which are not PE but satisfy $\Delta \notin \mc L_2$. \jo{Some guidelines on how to select the estimator parameters $d$ in \eqref{zdelay} as well as $\gamma_1$ and $\gamma_2$ in \eqref{inest} are given in Section~\ref{sec:nomtest}.}

\section{Numerical Verification on 1013-Machine ENTSO-E System: Nominal Test Case}
\label{sec:entsoe}
The performance of the proposed inertia DREM estimator is evaluated on the ENTSO-E system with topology and parameters as detailed in~\cite{ENTSOE}. The system has a total of $21382$ buses and $N=1013$ synchronous generators. Out of these, PFC units are connected at $N_\text{PFC}=871$ buses, while the generators at the remaining $N_\text{unc}=N-N_\text{PFC}$ buses have a constant active power setpoint. The AVR, governor, and PSS models of the PFC units are modeled each as detailed in~\cite{ENTSOE} and its references. The full detailed model consists of $133997$ differential-algebraic states.

The performance evaluation is undertaken as follows. At first, we demonstrate that the main frequency dynamics of the 1013-machine ENTSO-E system can indeed be captured by the model \eqref{sweq}. In addition, we show that for our selected benchmark system the PFC power injection $P_\text{PFC,tot}$ can be well-approximated using an aggregated, simple, model of the turbine-governor dynamics of the PFC units.

After this model validation step, we employ the DREM estimator \eqref{inest} to identify the overall inertia constant of the 1013-machine ENTSO-E system. 
The time-domain response of the system is obtained using the dynamic simulation software  RAMSES~\cite{ALC2016ja}.

\subsection{Aggregated Power System Model Including Turbine-Governor Dynamics of Primary-Controlled Units}
The majority of power plants in the considered ENTSO-E test system are thermal power plants. Therefore, similarly to \cite{weckesser17}, we assume that the turbine dynamics of the aggregated primary-controlled power plants can be represented by the TGOV1 model \cite{pourbeik13}
\begequ
\begin{split} 
	P_\text{PFC,tot}&=\frac{1+pT_z}{1+pT_p}\left(-K_P(\omega_\text{av}-\omega_0)\right),
\end{split}
\lab{pfc}
\endequ  
where $K_P\in\R_{>0}$ is the total primary (droop) control gain and $T_z\in\R_{>0}$ as well as $T_p\in\R_{>0}$ are time constants of the turbine-governor system of the aggregated generators\footnote{We recall that $p=\frac{d}{dt}$ is a differentiation operator.}. 

Thus, the overall aggregated power system dynamics are given by \eqref{sweq} and \eqref{pfc} and the corresponding system parameters for the aggregated ENTSO-E system are given in Table~\ref{tab:syspar}. 

\begrem
As indicated in \cite{weckesser17}, the model \eqref{pfc} has proven to be sufficiently accurate for representing PFC effects provided predominantly by steam power plants. If a significant amount of other units, such as hydro or gas power plants, also contribute to PFC, then the model \eqref{pfc} should be modified to account for these dynamics. Since we are mainly concerned with inertia estimation (and the dynamics \eqref{sweq} are independent of the PFC mix), we leave this extension for future research.
\endrem

\begin{table}
	\centering
	\caption{Aggregated ENTSO-E system parameters}
	\begin{tabular}{l | l}
		Parameter & Value \\\hline
		$S_B$ & 570.892 [GW]\\
		$H_\text{tot}$ & 3.665 [s]\\
		$\omega_0$ & 1.000 \jo{[pu]}\\
		$K_P$ & 2.495 [pu]\\
		$P_m$ & 0.498 [pu]\\
		$T_p$ & 12.983 [s]\\
		$T_z$ & 6.000 [s]\\
	\end{tabular}
	\label{tab:syspar}
\end{table}	

\subsection{Validation of Aggregated Model}
To validate the aggregated reduced-order model \eqref{sweq}, \eqref{pfc}, we first run a simulation of an outage of a power plant (the PFC unit 'FR918226' 
in France with $S_{B_{645}}=1755$ MW and $P_{m_{645}}=1455$ MW) in the bulk power system model using RAMSES. Then, we compute the aggregated data and parameters as defined in Section~\ref{sec:coi}. After that, we use the variables $P_{e,\text{tot}}$ and $P_{m,\text{tot}}$ as inputs to the aggregated model \eqref{sweq}, \eqref{pfc} and run a simulation of the aggregated model in Matlab/Simulink. Finally, the average frequency $f_\text{av}$ and total PFC injections $P_\text{PFC,tot}$ obtained with both models are compared. This comparison is illustrated in Fig.~\ref{fig:fav} and Fig.~\ref{fig:pfc}. 

These results show that the aggregated reduced-order model \eqref{sweq}, \eqref{pfc} offers a good approximation of the full-order ENTSO-E model. A small discrepancy at the frequency nadir can be explained by the fact that the non-PFC units are not explicitly considered in the model \eqref{sweq}, \eqref{pfc}. Therefore, we conclude that using the model \eqref{sweq} for the online-inertia estimator design is admissible. Likewise, the turbine-governor model \eqref{pfc} provides a good approximation of the PFC power injection $P_{\text{PFC,tot}}$.

\begin{figure}
	\centering
\subfloat[Comparison of the average electrical frequency $f_\text{av}$]
{
	\includegraphics[width=\linewidth]{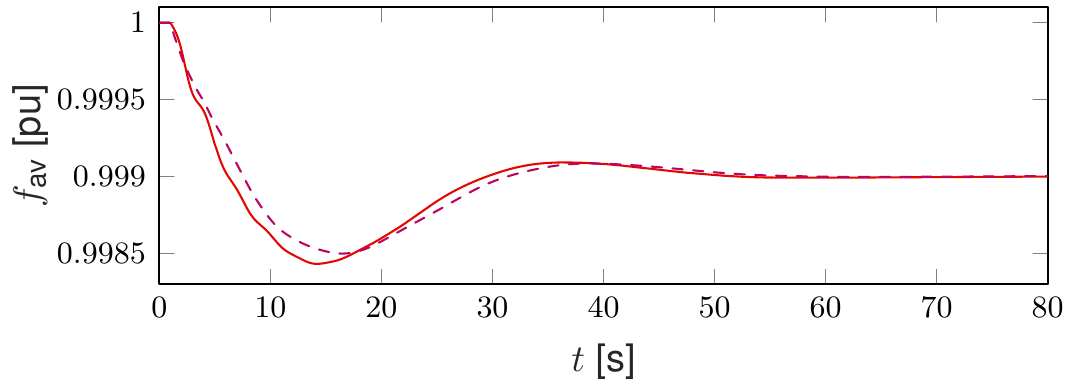}
	\lab{fig:fav}
}

\subfloat[Comparison of the PFC power injection $P_{\text{PFC,tot}}$]
{
~~~~~\includegraphics[width=.93\linewidth]{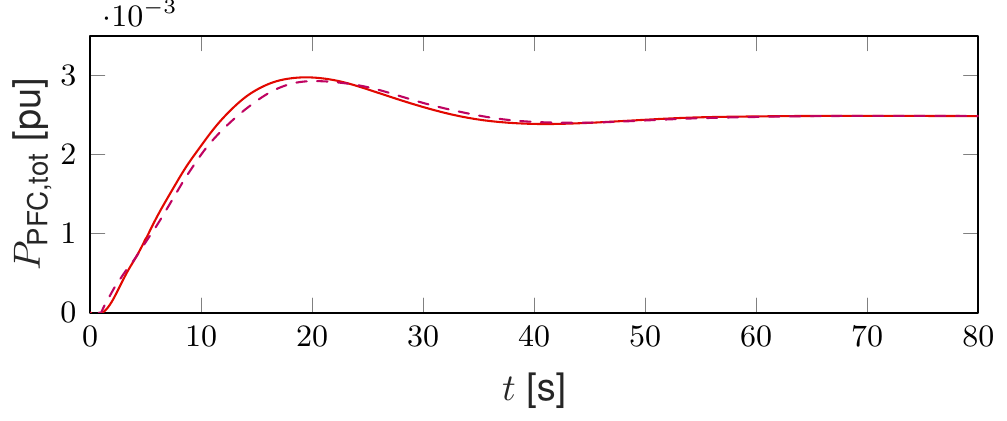}
	\lab{fig:pfc}
}
\caption{Comparison of the average electrical frequency $f_\text{av}$ and the PFC power injection $P_{\text{PFC,tot}}$ obtained from the bulk ENTSO-E system ('\textcolor[rgb]{0.88627,0.03137,0.00000}{--}') and from the aggregated model \eqref{sweq}, \eqref{pfc} ('\textcolor[rgb]{0.74902,0.01176,0.38039}{- -}').} 
\end{figure}

\subsection{Online-Inertia Estimation: Nominal Test Scenario}
\label{sec:nomtest}
The DREM-based estimator \eqref{inest}, see also Fig.~\ref{fig:drem_est}, is implemented in Matlab/Simulink. 
The estimator parameters are chosen as
\begequ
\alpha=10^3,\quad d=2,\quad \gamma_1=\gamma_2=10^{10}.
\notag 
\endequ 
This is motivated as follows. The operator used in \eqref{y_dot2} should not remove any important information from the filtered signals. Furthermore, the inertia constant mainly impacts the first few seconds of the power system's response to a disturbance. Hence, the choice $d=2$s. 
Finally, the values for the gains $\gamma_1$ and $\gamma_2$ have been tuned, such that the estimator possesses satisfactorily convergence properties for the ENTSO-E system under study.

\begin{figure}
	\centering
	\subfloat[Trajectories of the parameter estimates $\hat \eta$]
	{
	\includegraphics[width=\linewidth]{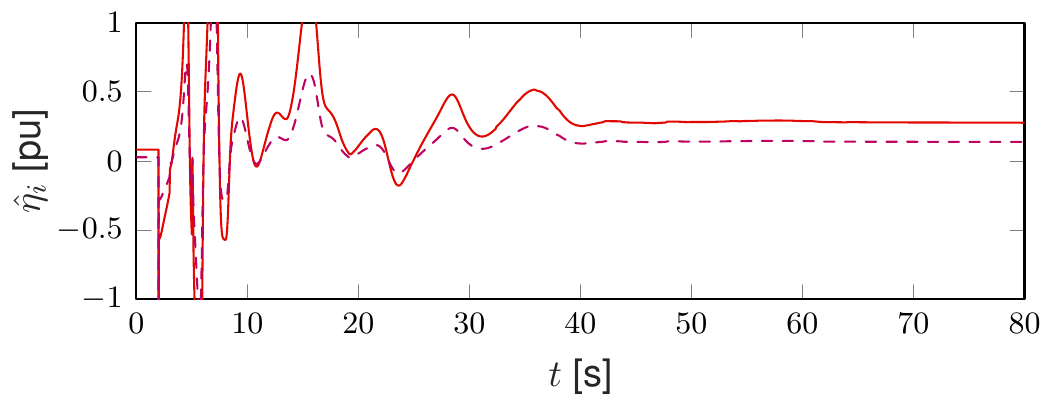}
	\label{fig:hateta} 
}

\subfloat[Trajectories of the parameter estimates $\hat \eta$ relative to the nominal parameter values $\eta$]
{
	\includegraphics[width=\linewidth]{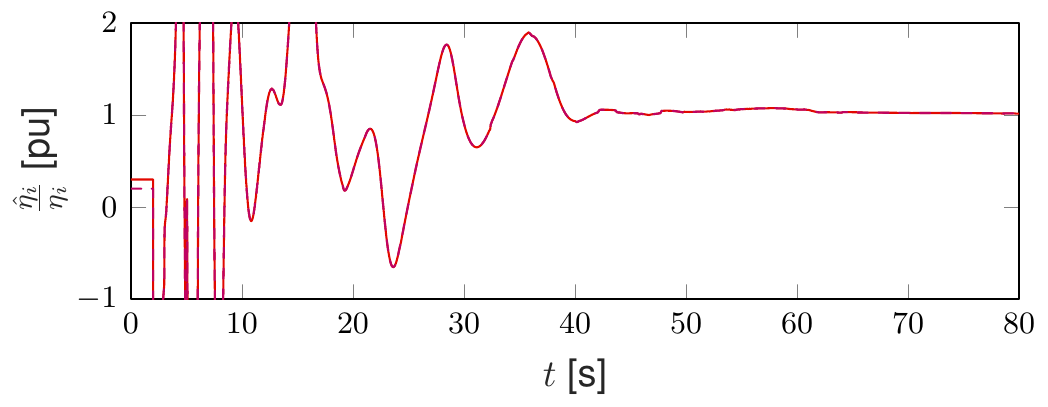}
	\label{fig:hatetadiveta} 
}
\caption{Absolute and relative trajectories of the parameter estimates $\hat \eta_i,$ respectively $\frac{\hat \eta_i}{\eta_i}$ ($i=1$ '\textcolor[rgb]{0.88627,0.03137,0.00000}{--}' and $i=2$ '\textcolor[rgb]{0.74902,0.01176,0.38039}{- -}') with initial condition $\hat \eta(0)=\diag(0.3,0.2)\eta$.}
\end{figure}

For the considered test system we work in per unit. Hence, the system base $S_B$ is removed from \eqref{sweq}. Moreover, we find from Table~\ref{tab:syspar} the nominal parameter values (see also \eqref{eta})
\begequ
\eta_1=\frac{1}{H_\text{tot}}=0.273,\quad \eta_2=\frac{P_{m,\text{PFC}}}{H_\text{tot}}=0.136.
\notag 
\endequ 
The performance of our proposed estimator is illustrated for the considered test case in Fig.~\ref{fig:hateta} with initial condition $\hat \eta(0)=\diag(0.3,0.2)\eta$. It can be observed that, after some initial transients, the estimates $\hat \eta$ converge to constant values.
This can also be appreciated from the trajectories in Fig.~\ref{fig:hatetadiveta}, which show the evolution of the relative errors $\frac{\hat\eta_i}{\eta_i}$, $i=1,2.$ 
The final estimates are
\begequ 
\hat \eta_1^s=0.290,\quad \hat \eta_2^s=0.145,
\notag 
\endequ 
or, expressed in relative terms,
\begequ 
\frac{\hat\eta_1^s}{\eta_1}=1.064,\quad \frac{\hat\eta_2^s}{\eta_2}=    1.064.
\notag 
\endequ 
Hence, the final estimation error is below $7\%$. In further numerical experiments we have observed a very similar behavior for a large variety of other initial conditions $\hat \eta(0)=\diag(\alpha_1,\alpha_2)\eta$ with $\alpha_i\in[0,30]$, $i=1,2$.

To assess the impact of the error introduced when using the model \eqref{pfc}, we take the measurement of $P_{\text{PFC,tot}}$ from the simulation and feed it directly as input to the estimator \eqref{inest} instead of using the model \eqref{pfc}. Doing so yields a stationary estimation error of only $1\%$. Hence, as one would expect, while the use of the model \eqref{pfc} reduces the number of required measurements, this is done at the expense of a (slightly) higher estimation error ($7\%$ instead of $1\%$). Furthermore, these experiments show that the approximations made in the derivation of the aggregated model \eqref{sweq} and the use of only data from PFC units (see Section~\ref{sec:model}) does not have significant effect on the estimation accuracy. 

\begin{figure}
	\includegraphics[width=\linewidth]{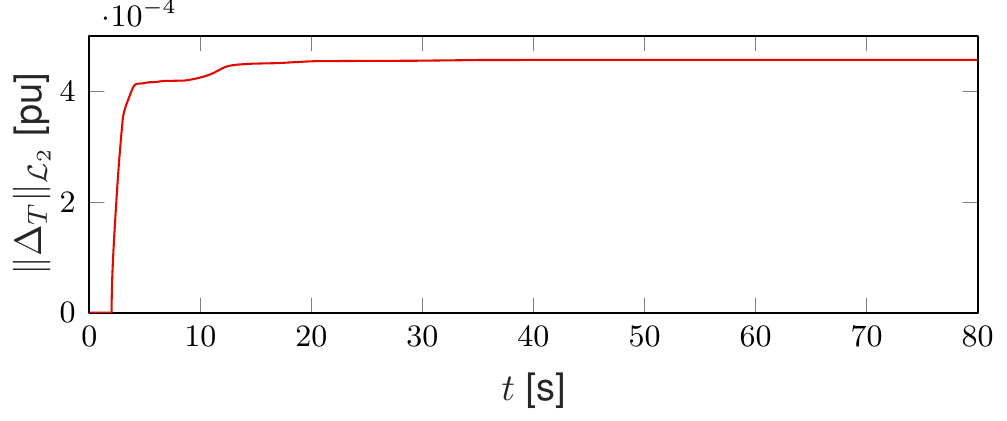}
	\caption{Evolution of the truncated $\mc L_2$-norm $\|\Delta_T\|_{\mc L_2}=\sqrt{\int_0^T \Delta^2 d\tau}$ of $\Delta$ defined in \eqref{Delta}.}
	\label{detPhi}
\end{figure}

\jo{With regard to tuning of the inertia estimator \eqref{inest}, we note the following.} As can be seen from \eqref{inesterr}, the fact that $\Delta\notin \mc L_2$ with $\Delta$  given in \eqref{Delta} is crucial for the performance of the inertia estimator \eqref{inest}. Since the numerical experiments are conducted on a finite time-horizon, we investigate the behavior of the truncated $\mc L_2$-norm of $\Delta$, {\em i.e.},
$$
\|\Delta_T\|_{\mc L_2}=\sqrt{\int_0^T \Delta^2 d\tau},
$$ 
instead of the $\mc L_2$-norm itself. The evolution of $\|\Delta_T\|_{\mc L_2}$ is plotted in Fig.~\ref{detPhi}. As one would expect, it increases significantly shortly after the disturbance and settles once the transients in $f_\text{av}$ and $P_\text{PFC,tot}$ have decayed. The magnitude of $\|\Delta_T\|_{\mc L_2}$ can be shaped by varying the magnitude of the delay $d$ in the DREM operator $[\mc H(\cdot)](t)=(\cdot)(t-d)$, see \eqref{opdelay}. In our experience, the best results can be obtained with $d\in[1,8]$s, which---as mentioned above---also roughly corresponds to the time span during which the inertia constant has the strongest influence on the aggregated power system trajectories, see also Fig.~\ref{fig:fav}. \jo{Once $d$ is fixed, the gains $\gamma_1$ and $\gamma_2$ have to be chosen large enough to ensure a desired convergence of the gradient algorithm \eqref{inest}. As in all parameter (or state) estimation problems the choice of $\gamma_1$ and $\gamma_2$ is a trade-off between speed of convergence and noise sensitivity.}

\section{Further Test Cases: Generator Outage and Rescheduling}
\label{sec:test}
The performance and accuracy of the inertia estimator \eqref{inest} is investigated via further test scenarios. The same ENTSO-E system and simulation software as well as the same estimator tuning gains as detailed in Section~\ref{sec:entsoe} are employed. The latter is crucial to assess whether a single set of tuning gains yields satisfactorily performances in diverse operating scenarios.
 
\subsection{Online-Inertia Estimation: Further Generator Outages}
To confirm the positive results of the previous section, we investigate the estimator performance in several further generator outage scenarios. \jo{We remark that for every considered outage scenario the inertia of the disconnected generator is removed from the total inertia constant $H_\text{tot}$. Hence, for each outage scenario the final total system inertia is different.}

In our second test scenario the PFC unit 'ES917736' in Spain with $S_{B_{644}}=1227$ MW and $P_{m_{644}}=1013$~MW is tripped. For this scenario we obtain a stationary estimation error of $7\%$, which is reduced to $2\%$ if the measurement of $P_{\text{PFC,tot}}$ is directly taken from the simulation in RAMSES as input to the estimator, instead of using the model \eqref{pfc}. Hence, despite the fact the generators are in different countries in each scenario, the obtained results are very similar to those of the previous scenario investigated in Section~\ref{sec:entsoe}. 

\jo{To further evaluate the dependency of the estimation accuracy on the geographical location and size of the generator outage, we investigate 23 further generator outage scenarios across the whole ENTSO-E area. To this end, we trip randomly selected generators in Bulgaria (BG), Germany (DE), France (FR), Italy (IT), Romania (RO), Serbia (RS), Spain (ES) and Turkey (TR). Thereby, the disturbance sizes in terms of lost generation range from 500~MW to 1200 MW 
}

\begin{figure}
	\centering 
	\includegraphics[width=\linewidth]{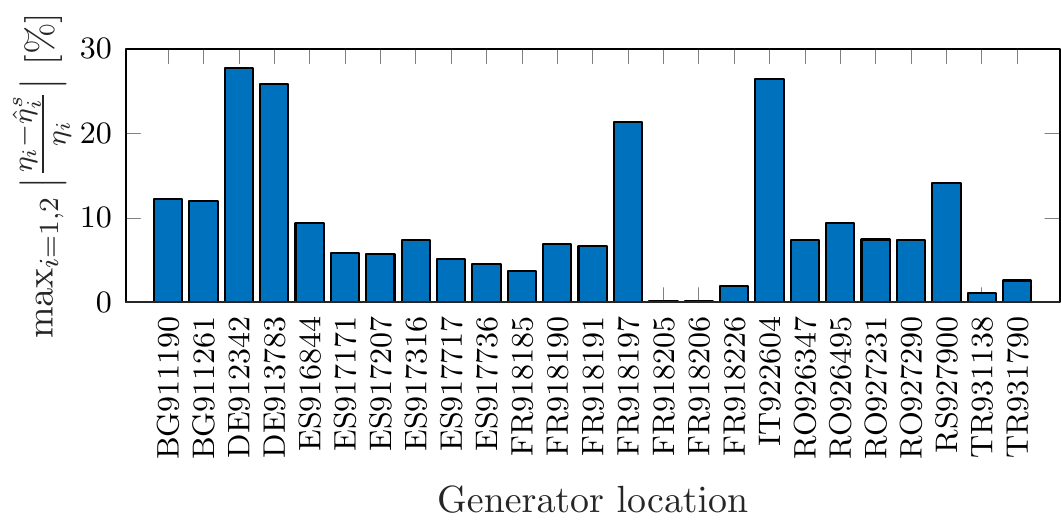}
	\caption{\jo{Relative estimation errors for 25 generator outage scenarios across the whole ENTSO-E area.}}
	\label{fig:est_results} 
\end{figure}

\jo{We find that for 21 out of the 25 outage scenarios considered in total, the estimation error is $15\%$ or lower and for some cases it is even below $1\%,$ see Fig.~\ref{fig:est_results}. Only for 4 cases, we obtain an estimation error larger than $15\%$. These cases correspond to disturbances with a magnitude above 800 MW in Germany, France and Italy. 
This suggests that neither the location nor the size of the disturbance are the key decisive factors for the estimator performance. 
}	

\jo{In order to further investigate these findings and, more importantly, the validity of the obtained estimated total inertia coefficient $\hat H_\text{tot}=\frac{1}{\hat \eta_1},$ we simulate the average frequency dynamics \eqref{sweq} for two settings: First, by using the nominal inertia coefficient $H_\text{tot}$ obtained directly from the data in \cite{ENTSOE} (and references therein) and, second, by using the estimated inertia coefficient $\hat H_\text{tot}.$ In both cases, the time series for the signals $P_{m,\text{PFC}}$ and $P_{e,\text{PFC}}$ in \eqref{sweq} are taken from the simulation results in RAMSES, while $P_\text{PFC,tot}$ is modeled with \eqref{pfc}. The obtained frequency curves are denoted by $\omega_{\text{av}}^{H_\text{tot}}$ and $\omega_{\text{av}}^{\hat H_\text{tot}}.$ The average frequency obtained from the full-system simulation in RAMSES is denoted by $\omega_{\text{av}}^\text{RAMSES}.$
}

\begin{figure}
	\centering 
	\includegraphics[width=\linewidth]{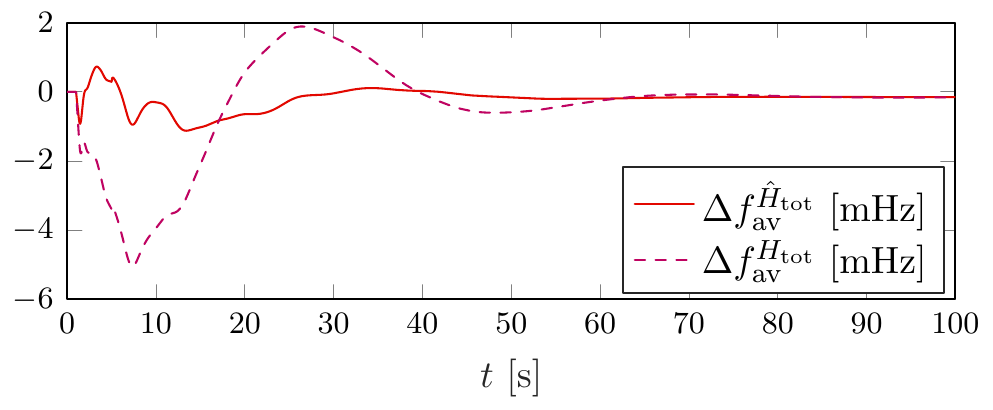}
	\caption{\jo{Trajectories of $\Delta f_{\text{av}}^{\hat H_\text{tot}}$ and $\Delta f_{\text{av}}^{H_\text{tot}}$ defined in \eqref{deltaomega} for the outage of the unit 'DE912342' in Germany with $S_{B_{645}}=2154$ MW and $P_{m_ {154}}=1078.5$ MW.}}
	\label{fig:comp_f} 
\end{figure}

\jo{For the outage of the unit 'DE912342' in Germany with $S_{B_{645}}=2154$ MW and $P_{m_ {154}}=1078.5$ MW and a relative estimation error in $\eta_1$ of $27.5\%,$ the evolution of the frequency deviations
\begequ
\begin{split}
\Delta f_{\text{av}}^{\hat H_\text{tot}}&=\frac{1}{2\pi}\left(\omega_{\text{av}}^{\hat H_\text{tot}}-\omega_{\text{av}}^\text{RAMSES}\right),\\
\Delta f_{\text{av}}^{H_\text{tot}}&=\frac{1}{2\pi}\left(\omega_{\text{av}}^{H_\text{tot}}-\omega_{\text{av}}^\text{RAMSES}\right),
\end{split}
\label{deltaomega}
\endequ 
are shown in Fig.~\ref{fig:comp_f}. Clearly, the evolution of $f_{\text{av}}^{\hat H_\text{tot}}$ resembles very closely that of $f_{\text{av}}^\text{RAMSES},$ while the evolution of $f_{\text{av}}^{H_\text{tot}}$ shows some larger discrepancies with respect to $f_{\text{av}}^\text{RAMSES}$ (though for both cases the deviations are in the mHz-range). This may indicate that---at least with the model \eqref{sweq}, \eqref{pfc}---the estimated inertia coefficient $\hat H_\text{tot}$ provides a more accurate representation of the true system response and, hence, of the effective inertia than the nominal inertia coefficient $H_\text{tot}$ calculated directly from the system data.
}

\begin{figure}
	\centering 
	\includegraphics[width=1.\linewidth]{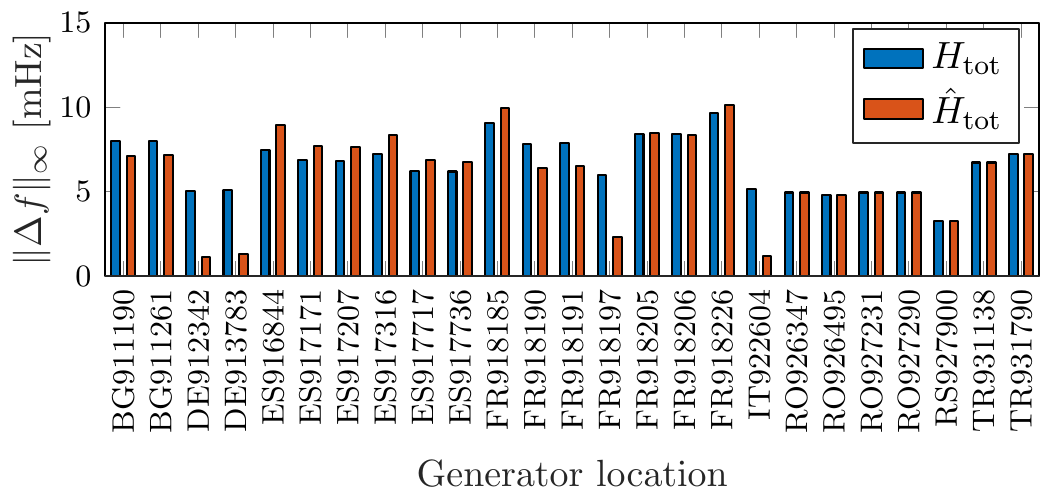}
	\caption{\jo{Maximum frequency deviations $\|\Delta f_{\text{av}}^{\hat H_\text{tot}}\|_\infty$ and $\|\Delta f_{\text{av}}^{H_\text{tot}}\|_\infty$ in mHz with respect to average frequency obtained from full-system simulations in RAMSES for 25 generator outage scenarios across the whole ENTSO-E area.}}
	\label{fig:est_results_deltaf} 
\end{figure}

\jo{The same experiment is performed for all 24 other outage scenarios and the maximum frequency errors are shown in Fig.~\ref{fig:est_results_deltaf}. There is a clear trend that whenever the estimation error for $\eta$ is above $15\%$, then 
$$
\|\Delta f_{\text{av}}^{\hat H_\text{tot}}\|_\infty<\|\Delta f_{\text{av}}^{H_\text{tot}}\|_\infty,
$$ 
where $\|\cdot\|$ denotes the vector infinity norm, {\em i.e.,} the estimated inertia coefficient $\hat H_\text{tot}$ provides a better characterization of the actual system behavior than the nominal one $H_\text{tot},$ at least with the model \eqref{sweq}, \eqref{pfc}.
}

\jo{This observation opens many new, interesting questions regarding the influence that dynamic phenomena associated to voltage, reactive power or renewable generation and load dynamics may have on the behavior of the system frequency. Similar observations on the load voltage dynamics affecting the effective inertia of the system were made in \cite{orum15,weckesser17}. Given the complexity of the employed ENTSO-E model, we leave a detailed investigation of these aspects as well as a possible extension of the model \eqref{sweq}, \eqref{pfc} to incorporate some of them for future work.
}

\subsection{Online-Inertia Estimation: Rescheduling Events}
As discussed in Section~\ref{sec:entsoe}, a certain level of variation of the measurement signal(s), also referred to as excitation in the parameter identification literature \cite{sastry11,astolfi07,narendra12}, is essential for the estimation problem to be feasible. 
The frequency variation under usual operating conditions does---in our experience---not possess a sufficient level of excitation. Therefore, thus far and in line with other online inertia estimation approaches \cite{wall12,wall14}, we have focused our performance analysis on outage scenarios. While these are clear opportunities for inertia estimation, their occurrence is rather infrequent and unscheduled. Hence, the question arises whether there are other operating scenarios, which can be exploited to perform the estimation.
In this context, imbalances resulting from scheduling changes can lead to significant frequency variations \cite{weissbach09,weissbach09_2,hirth15}. In particular, this applies to rescheduling events at full hours \cite{weissbach09,weissbach09_2,hirth15}. Hence, these are frequent, recurring, and scheduled frequency variations, which can be another useful source for inertia estimation. 

\begin{figure}
	\centering 
	\includegraphics[width=\linewidth]{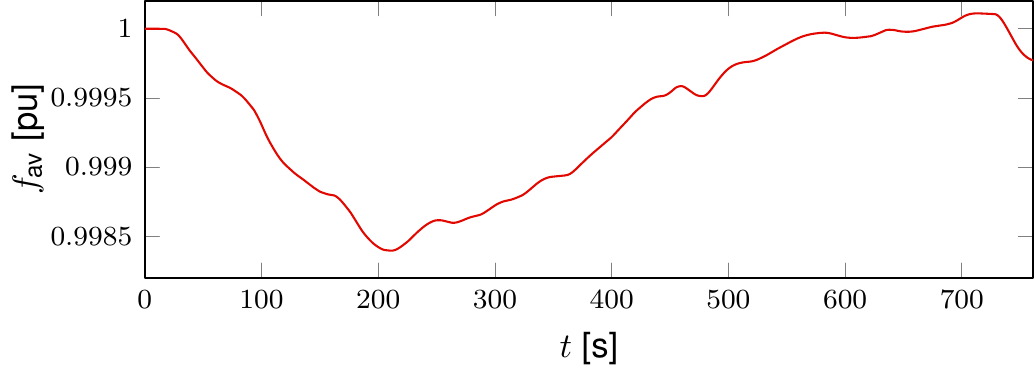}
	\caption{Typical evolution of the (average) system frequency during a rescheduling process, based on \cite[Fig. 1.2]{weissbach09_2}.}
	\label{fig:fav_sched} 
\end{figure}

\begin{figure}
	\centering 
	\includegraphics[width=\linewidth]{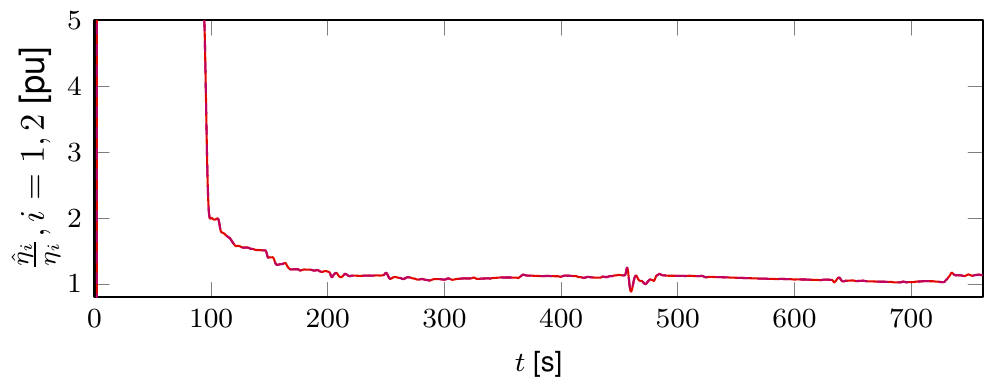}

\caption{Trajectories of the parameter estimates $\hat \eta$ relative to the nominal parameter values $\eta$, {\em i.e.,} $\frac{\hat \eta_i}{\eta_i}$ ($i=1$ '\textcolor[rgb]{0.88627,0.03137,0.00000}{--}' and $i=2$ '\textcolor[rgb]{0.74902,0.01176,0.38039}{- -}') with initial condition $\hat \eta(0)=\diag(0.3,0.2)\eta$ in the rescheduling scenario.}
	\label{fig:hatetadiveta_sched} 
\end{figure}

A typical frequency evolution during a rescheduling process is shown in Fig.~\ref{fig:fav_sched}. This exemplary frequency trajectory $f_\text{av}$ is based on \cite[Fig. 1.2]{weissbach09_2} and has been reproduced in RAMSES using the ENTSO-E system described in Section~\ref{sec:entsoe} and performing scheduled power setpoint changes to generators and loads. Furthermore, we measure $P_{\text{PFC,tot}}$ directly from the simulation results in RAMSES. 
Feeding both signals $f_\text{av}$ and $P_{\text{PFC,tot}}$ to the estimator \eqref{inest} yields the relative estimation trajectories $\frac{\hat \eta_i}{\eta_i}$, $i=1,2,$ shown in Fig.~\ref{fig:hatetadiveta_sched}. It can be seen that the trajectories converge to a band around the nominal value of $1$. 
We find that the maximum average relative estimation error $e_{\text{avg}}$ over the time window $t\in[T_1,T_2]=[300,761]$s is given by
\begequ
e_{\text{avg}}=\max_{i=1,2}\left(\frac{1}{T_2-T_1}\int_{T_1}^{T_2} \left | \frac{\eta_i-\hat \eta_i(\tau)}{\eta_i}\right| d\tau\right) =0.08.
\label{eavg}
\endequ
Hence, also in this scenario, the estimation error is below $10\%.$ This confirms both that rescheduling events can provide useful data for inertia estimation and that the proposed DREM-based estimator \eqref{inest} is well-suited for this task.

\begrem
In the rescheduling scenario the variations in the generator power injection are not solely dictated by the model \eqref{pfc}, but also by the rescheduling sequences, {\em i.e.},
\begequ 
\Delta P=	P_{m,\text{PFC}}+P_\text{PFC,tot}+P_\text{res,PFC}-P_{e,\text{PFC}},
\notag 
\endequ 
where $P_\text{res,PFC}:\R_{\geq0}\to\R_{\geq0}$ denotes the power variation of the PFC units due to the rescheduling event.
Therefore in this scenario using the model \eqref{pfc} does not yield any significant advantages and is thus omitted.
\endrem

\section{Conclusions}
\label{sec:con}
An algorithm to monitor in real time the inertia constant of a large-scale power system has been presented. The increasing penetration of renewable energy units makes this a highly desirable feature to gain a better understanding of the system's inertial frequency response and the security of the system in near to real time. In addition to the inertia constant, the aggregated mechanical power setpoint of the PFC generators is also estimated. 

The proposed estimator is based on a nonlinear, aggregated power system model and constructed using the recently proposed DREM procedure. Its performance has been demonstrated \jo{via 25 test scenarios, in 21 of which the estimation error compared to the COI inertia constant was below $15\%$ while in all of them the response of the aggregated system based on the estimated inertia matches the simulated one with an error of only a few mHz.} Remarkably, our approach is also applicable in rescheduling events, which occur numerous times every day in any deregulated power system. This is a distinguished feature compared to other existing solutions and significantly enhances the applicability of our solution.

The proposed estimator opens the door for many subsequent applications in the realm of power system protection and real-time control. Exploring these possibilities will be part of our future research. \jo{
In addition, we plan to investigate the impact of both measurement data resolution and noise on the estimation performance.}

\jo{\section*{Acknowledgment}
The authors  would like to thank Prof. Thierry Van Cutsem for many helpful comments on the topics of this paper.}

\bibliographystyle{IEEEtran}
\bibliography{bib_microgrids}

\begin{IEEEbiography}
[{\includegraphics[width=1in,height=1.25in,clip,keepaspectratio]{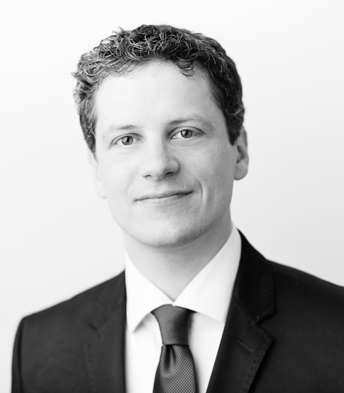}}]
{Johannes Schiffer} received the Diploma degree in engineering cybernetics from the University of Stuttgart, Germany, in 2009 and the Ph.D. degree (Dr.-Ing.) in electrical engineering from Technische Universit\"at (TU) Berlin, Germany, in 2015.

He currently holds the chair of Control Systems and Network Control Technology at Brandenburgische Technische Universit\"at Cottbus-Senftenberg, Germany. Prior to that, he has held appointments as Lecturer (Assistant Professor) at the School of Electronic and Electrical Engineering, University of Leeds, U.K. and as Research Associate in the Control Systems Group and at the Chair of Sustainable Electric Networks and Sources of Energy both at TU Berlin. 

In 2017 he and his co-workers received the Automatica Paper Prize over the years 2014-2016. His current research interests include distributed control and analysis of complex networks with application to microgrids and power systems.
\end{IEEEbiography}

\begin{IEEEbiography}
[{\includegraphics[width=1in,height=1.25in,clip,keepaspectratio]{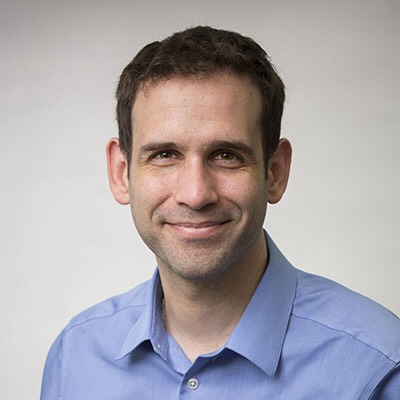}}]
{Petros Aristidou} (S'10-M'15) received a Diploma in Electrical and Computer Engineering from the National Technical University of Athens, Greece, in 2010, and a PhD in Engineering Sciences from the University of Li\`{e}ge, Belgium, in 2015. He is currently a Lecturer (Assistant Professor) in Smart Energy Systems at the University of Leeds, U.K. Prior to that, he was a Postdoctoral Researcher at the Power System Laboratory at ETH Zurich. 

His current research interests include power system dynamics and control, and developing numerical methods for analysing large-scale dynamic networks.
\end{IEEEbiography}

\begin{IEEEbiography}
[{\includegraphics[width=1in,height=1.25in,clip,keepaspectratio]{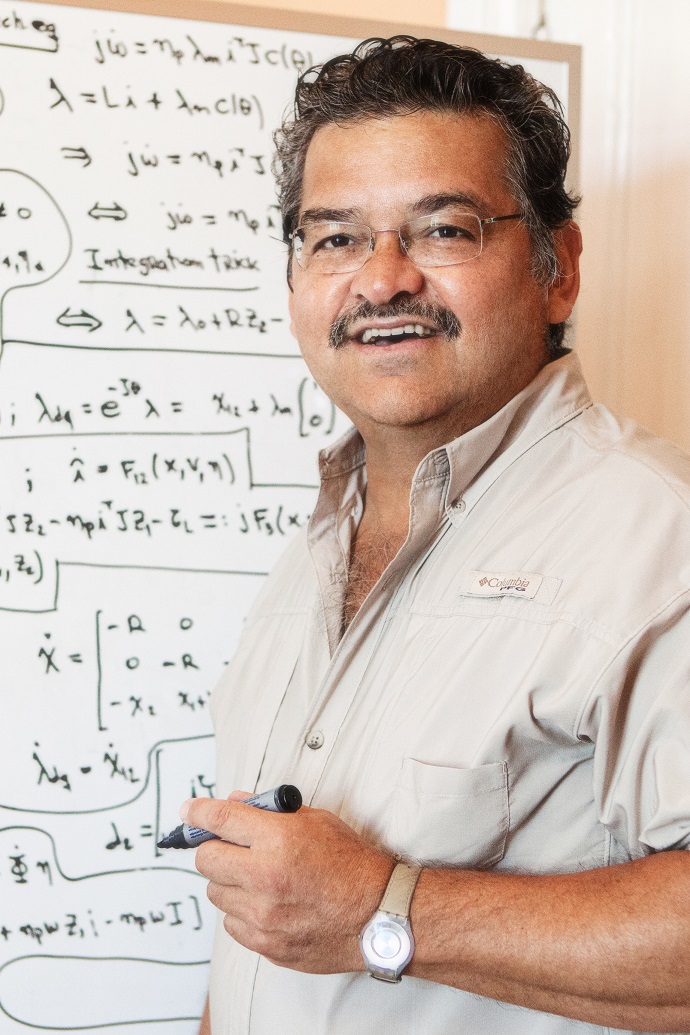}}]
{Romeo Ortega}
(S'81, M'85, SM'98, F'99) was born in Mexico. He obtained his BSc in
Electrical and Mechanical Engineering from the National University of
Mexico, Master of Engineering from Polytechnical Institute of
Leningrad, USSR, and the Docteur D`Etat from the Polytechnical
Institute of Grenoble, France in 1974, 1978 and 1984 respectively.

He then joined the National University of Mexico, where he worked until
1989. He was a Visiting Professor at the University of Illinois in
1987-88 and at the McGill University in 1991-1992, and a Fellow of the
Japan Society for Promotion of Science in 1990-1991.
He has been a member of the French National Researcher Council (CNRS) since
June 1992.  Currently he is in the Laboratoire de Signaux et Systemes (SUPELEC)
 in Gif--sur--Yvette.  His research interests are in the fields of
nonlinear and adaptive control, with special emphasis on applications. 

Dr Ortega has published  three books and more than 290 scientific papers in international journals, with an h-index of 79. He
has supervised more than 30 PhD thesis. He  has served as chairman in several IFAC and IEEE committees and participated in various editorial boards.
\end{IEEEbiography}

\end{document}